\documentclass[12pt]{article}
\usepackage{latexsym}
\usepackage{amsfonts}
\title{Progress in study of ${\cal N}=4$ SYM effective action}
\author{I.L. Buchbinder}

\date{{\it Department of Theoretical Physics, \\Tomsk State Pedagogical
University, \\634041, Russia}}

\begin{document}
\immediate\write16{<<WARNING: LINEDRAW macros work with emTeX-dvivers
                    and other drivers supporting emTeX \special's
                    (dviscr, dvihplj, dvidot, dvips, dviwin, etc.) >>}
\newdimen\Lengthunit       \Lengthunit  = 1.5cm
\newcount\Nhalfperiods     \Nhalfperiods= 9
\newcount\magnitude        \magnitude = 1000

\catcode`\*=11
\newdimen\L*   \newdimen\d*   \newdimen\d**
\newdimen\dm*  \newdimen\dd*  \newdimen\dt*
\newdimen\a*   \newdimen\b*   \newdimen\c*
\newdimen\a**  \newdimen\b**
\newdimen\xL*  \newdimen\yL*
\newdimen\rx*  \newdimen\ry*
\newdimen\tmp* \newdimen\linwid*

\newcount\k*   \newcount\l*   \newcount\m*
\newcount\k**  \newcount\l**  \newcount\m**
\newcount\n*   \newcount\dn*  \newcount\r*
\newcount\N*   \newcount\*one \newcount\*two  \*one=1 \*two=2
\newcount\*ths \*ths=1000
\newcount\angle*  \newcount\q*  \newcount\q**
\newcount\angle** \angle**=0
\newcount\sc*     \sc*=0

\newtoks\cos*  \cos*={1}
\newtoks\sin*  \sin*={0}

\catcode`\[=13

\def\rotate(#1){\advance\angle**#1\angle*=\angle**
\q**=\angle*\ifnum\q**<0\q**=-\q**\fi
\ifnum\q**>360\q*=\angle*\divide\q*360\multiply\q*360\advance\angle*-\q*\fi
\ifnum\angle*<0\advance\angle*360\fi\q**=\angle*\divide\q**90\q**=\q**
\def\sgcos*{+}\def\sgsin*{+}\relax
\ifcase\q**\or
 \def\sgcos*{-}\def\sgsin*{+}\or
 \def\sgcos*{-}\def\sgsin*{-}\or
 \def\sgcos*{+}\def\sgsin*{-}\else\fi
\q*=\q**
\multiply\q*90\advance\angle*-\q*
\ifnum\angle*>45\sc*=1\angle*=-\angle*\advance\angle*90\else\sc*=0\fi
\def[##1,##2]{\ifnum\sc*=0\relax
\edef\cs*{\sgcos*.##1}\edef\sn*{\sgsin*.##2}\ifcase\q**\or
 \edef\cs*{\sgcos*.##2}\edef\sn*{\sgsin*.##1}\or
 \edef\cs*{\sgcos*.##1}\edef\sn*{\sgsin*.##2}\or
 \edef\cs*{\sgcos*.##2}\edef\sn*{\sgsin*.##1}\else\fi\else
\edef\cs*{\sgcos*.##2}\edef\sn*{\sgsin*.##1}\ifcase\q**\or
 \edef\cs*{\sgcos*.##1}\edef\sn*{\sgsin*.##2}\or
 \edef\cs*{\sgcos*.##2}\edef\sn*{\sgsin*.##1}\or
 \edef\cs*{\sgcos*.##1}\edef\sn*{\sgsin*.##2}\else\fi\fi
\cos*={\cs*}\sin*={\sn*}\global\edef\gcos*{\cs*}\global\edef\gsin*{\sn*}}\relax
\ifcase\angle*[9999,0]\or
[999,017]\or[999,034]\or[998,052]\or[997,069]\or[996,087]\or
[994,104]\or[992,121]\or[990,139]\or[987,156]\or[984,173]\or
[981,190]\or[978,207]\or[974,224]\or[970,241]\or[965,258]\or
[961,275]\or[956,292]\or[951,309]\or[945,325]\or[939,342]\or
[933,358]\or[927,374]\or[920,390]\or[913,406]\or[906,422]\or
[898,438]\or[891,453]\or[882,469]\or[874,484]\or[866,499]\or
[857,515]\or[848,529]\or[838,544]\or[829,559]\or[819,573]\or
[809,587]\or[798,601]\or[788,615]\or[777,629]\or[766,642]\or
[754,656]\or[743,669]\or[731,681]\or[719,694]\or[707,707]\or
\else[9999,0]\fi}

\catcode`\[=12

\def\GRAPH(hsize=#1)#2{\hbox to #1\Lengthunit{#2\hss}}

\def\Linewidth#1{\global\linwid*=#1\relax
\global\divide\linwid*10\global\multiply\linwid*\mag
\global\divide\linwid*100\special{em:linewidth \the\linwid*}}

\Linewidth{.4pt}
\def\sm*{\special{em:moveto}}
\def\sl*{\special{em:lineto}}
\let\moveto=\sm*
\let\lineto=\sl*
\newbox\spm*   \newbox\spl*
\setbox\spm*\hbox{\sm*}
\setbox\spl*\hbox{\sl*}

\def\mov#1(#2,#3)#4{\rlap{\L*=#1\Lengthunit
\xL*=#2\L* \yL*=#3\L*
\xL*=\xscale\xL* \yL*=\yscale\yL*
\rx* \the\cos*\xL* \tmp* \the\sin*\yL* \advance\rx*-\tmp*
\ry* \the\cos*\yL* \tmp* \the\sin*\xL* \advance\ry*\tmp*
\kern\rx*\raise\ry*\hbox{#4}}}

\def\rmov*(#1,#2)#3{\rlap{\xL*=#1\yL*=#2\relax
\rx* \the\cos*\xL* \tmp* \the\sin*\yL* \advance\rx*-\tmp*
\ry* \the\cos*\yL* \tmp* \the\sin*\xL* \advance\ry*\tmp*
\kern\rx*\raise\ry*\hbox{#3}}}

\def\lin#1(#2,#3){\rlap{\sm*\mov#1(#2,#3){\sl*}}}

\def\arr*(#1,#2,#3){\rmov*(#1\dd*,#1\dt*){\sm*
\rmov*(#2\dd*,#2\dt*){\rmov*(#3\dt*,-#3\dd*){\sl*}}\sm*
\rmov*(#2\dd*,#2\dt*){\rmov*(-#3\dt*,#3\dd*){\sl*}}}}

\def\arrow#1(#2,#3){\rlap{\lin#1(#2,#3)\mov#1(#2,#3){\relax
\d**=-.012\Lengthunit\dd*=#2\d**\dt*=#3\d**
\arr*(1,10,4)\arr*(3,8,4)\arr*(4.8,4.2,3)}}}

\def\arrlin#1(#2,#3){\rlap{\L*=#1\Lengthunit\L*=.5\L*
\lin#1(#2,#3)\rmov*(#2\L*,#3\L*){\arrow.1(#2,#3)}}}

\def\dasharrow#1(#2,#3){\rlap{{\Lengthunit=0.9\Lengthunit
\dashlin#1(#2,#3)\mov#1(#2,#3){\sm*}}\mov#1(#2,#3){\sl*
\d**=-.012\Lengthunit\dd*=#2\d**\dt*=#3\d**
\arr*(1,10,4)\arr*(3,8,4)\arr*(4.8,4.2,3)}}}

\def\clap#1{\hbox to 0pt{\hss #1\hss}}

\def\ind(#1,#2)#3{\rlap{\L*=.1\Lengthunit
\xL*=#1\L* \yL*=#2\L*
\rx* \the\cos*\xL* \tmp* \the\sin*\yL* \advance\rx*-\tmp*
\ry* \the\cos*\yL* \tmp* \the\sin*\xL* \advance\ry*\tmp*
\kern\rx*\raise\ry*\hbox{\lower2pt\clap{$#3$}}}}

\def\sh*(#1,#2)#3{\rlap{\dm*=\the\n*\d**
\xL*=\xscale\dm* \yL*=\yscale\dm* \xL*=#1\xL* \yL*=#2\yL*
\rx* \the\cos*\xL* \tmp* \the\sin*\yL* \advance\rx*-\tmp*
\ry* \the\cos*\yL* \tmp* \the\sin*\xL* \advance\ry*\tmp*
\kern\rx*\raise\ry*\hbox{#3}}}

\def\calcnum*#1(#2,#3){\a*=1000sp\b*=1000sp\a*=#2\a*\b*=#3\b*
\ifdim\a*<0pt\a*-\a*\fi\ifdim\b*<0pt\b*-\b*\fi
\ifdim\a*>\b*\c*=.96\a*\advance\c*.4\b*
\else\c*=.96\b*\advance\c*.4\a*\fi
\k*\a*\multiply\k*\k*\l*\b*\multiply\l*\l*
\m*\k*\advance\m*\l*\n*\c*\r*\n*\multiply\n*\n*
\dn*\m*\advance\dn*-\n*\divide\dn*2\divide\dn*\r*
\advance\r*\dn*
\c*=\the\Nhalfperiods5sp\c*=#1\c*\ifdim\c*<0pt\c*-\c*\fi
\multiply\c*\r*\N*\c*\divide\N*10000}

\def\dashlin#1(#2,#3){\rlap{\calcnum*#1(#2,#3)\relax
\d**=#1\Lengthunit\ifdim\d**<0pt\d**-\d**\fi
\divide\N*2\multiply\N*2\advance\N*\*one
\divide\d**\N*\sm*\n*\*one\sh*(#2,#3){\sl*}\loop
\advance\n*\*one\sh*(#2,#3){\sm*}\advance\n*\*one
\sh*(#2,#3){\sl*}\ifnum\n*<\N*\repeat}}

\def\dashdotlin#1(#2,#3){\rlap{\calcnum*#1(#2,#3)\relax
\d**=#1\Lengthunit\ifdim\d**<0pt\d**-\d**\fi
\divide\N*2\multiply\N*2\advance\N*1\multiply\N*2\relax
\divide\d**\N*\sm*\n*\*two\sh*(#2,#3){\sl*}\loop
\advance\n*\*one\sh*(#2,#3){\kern-1.48pt\lower.5pt\hbox{\rm.}}\relax
\advance\n*\*one\sh*(#2,#3){\sm*}\advance\n*\*two
\sh*(#2,#3){\sl*}\ifnum\n*<\N*\repeat}}

\def\shl*(#1,#2)#3{\kern#1#3\lower#2#3\hbox{\unhcopy\spl*}}

\def\trianglin#1(#2,#3){\rlap{\toks0={#2}\toks1={#3}\calcnum*#1(#2,#3)\relax
\dd*=.57\Lengthunit\dd*=#1\dd*\divide\dd*\N*
\divide\dd*\*ths \multiply\dd*\magnitude
\d**=#1\Lengthunit\ifdim\d**<0pt\d**-\d**\fi
\multiply\N*2\divide\d**\N*\sm*\n*\*one\loop
\shl**{\dd*}\dd*-\dd*\advance\n*2\relax
\ifnum\n*<\N*\repeat\n*\N*\shl**{0pt}}}

\def\wavelin#1(#2,#3){\rlap{\toks0={#2}\toks1={#3}\calcnum*#1(#2,#3)\relax
\dd*=.23\Lengthunit\dd*=#1\dd*\divide\dd*\N*
\divide\dd*\*ths \multiply\dd*\magnitude
\d**=#1\Lengthunit\ifdim\d**<0pt\d**-\d**\fi
\multiply\N*4\divide\d**\N*\sm*\n*\*one\loop
\shl**{\dd*}\dt*=1.3\dd*\advance\n*\*one
\shl**{\dt*}\advance\n*\*one
\shl**{\dd*}\advance\n*\*two
\dd*-\dd*\ifnum\n*<\N*\repeat\n*\N*\shl**{0pt}}}

\def\w*lin(#1,#2){\rlap{\toks0={#1}\toks1={#2}\d**=\Lengthunit\dd*=-.12\d**
\divide\dd*\*ths \multiply\dd*\magnitude
\N*8\divide\d**\N*\sm*\n*\*one\loop
\shl**{\dd*}\dt*=1.3\dd*\advance\n*\*one
\shl**{\dt*}\advance\n*\*one
\shl**{\dd*}\advance\n*\*one
\shl**{0pt}\dd*-\dd*\advance\n*1\ifnum\n*<\N*\repeat}}

\def\l*arc(#1,#2)[#3][#4]{\rlap{\toks0={#1}\toks1={#2}\d**=\Lengthunit
\dd*=#3.037\d**\dd*=#4\dd*\dt*=#3.049\d**\dt*=#4\dt*\ifdim\d**>10mm\relax
\d**=.25\d**\n*\*one\shl**{-\dd*}\n*\*two\shl**{-\dt*}\n*3\relax
\shl**{-\dd*}\n*4\relax\shl**{0pt}\else
\ifdim\d**>5mm\d**=.5\d**\n*\*one\shl**{-\dt*}\n*\*two
\shl**{0pt}\else\n*\*one\shl**{0pt}\fi\fi}}

\def\d*arc(#1,#2)[#3][#4]{\rlap{\toks0={#1}\toks1={#2}\d**=\Lengthunit
\dd*=#3.037\d**\dd*=#4\dd*\d**=.25\d**\sm*\n*\*one\shl**{-\dd*}\relax
\n*3\relax\sh*(#1,#2){\xL*=\xscale\dd*\yL*=\yscale\dd*
\kern#2\xL*\lower#1\yL*\hbox{\sm*}}\n*4\relax\shl**{0pt}}}

\def\shl**#1{\c*=\the\n*\d**\d*=#1\relax
\a*=\the\toks0\c*\b*=\the\toks1\d*\advance\a*-\b*
\b*=\the\toks1\c*\d*=\the\toks0\d*\advance\b*\d*
\a*=\xscale\a*\b*=\yscale\b*
\rx* \the\cos*\a* \tmp* \the\sin*\b* \advance\rx*-\tmp*
\ry* \the\cos*\b* \tmp* \the\sin*\a* \advance\ry*\tmp*
\raise\ry*\rlap{\kern\rx*\unhcopy\spl*}}

\def\wlin*#1(#2,#3)[#4]{\rlap{\toks0={#2}\toks1={#3}\relax
\c*=#1\l*\c*\c*=.01\Lengthunit\m*\c*\divide\l*\m*
\c*=\the\Nhalfperiods5sp\multiply\c*\l*\N*\c*\divide\N*\*ths
\divide\N*2\multiply\N*2\advance\N*\*one
\dd*=.002\Lengthunit\dd*=#4\dd*\multiply\dd*\l*\divide\dd*\N*
\divide\dd*\*ths \multiply\dd*\magnitude
\d**=#1\multiply\N*4\divide\d**\N*\sm*\n*\*one\loop
\shl**{\dd*}\dt*=1.3\dd*\advance\n*\*one
\shl**{\dt*}\advance\n*\*one
\shl**{\dd*}\advance\n*\*two
\dd*-\dd*\ifnum\n*<\N*\repeat\n*\N*\shl**{0pt}}}

\def\wavebox#1{\setbox0\hbox{#1}\relax
\a*=\wd0\advance\a*14pt\b*=\ht0\advance\b*\dp0\advance\b*14pt\relax
\hbox{\kern9pt\relax
\rmov*(0pt,\ht0){\rmov*(-7pt,7pt){\wlin*\a*(1,0)[+]\wlin*\b*(0,-1)[-]}}\relax
\rmov*(\wd0,-\dp0){\rmov*(7pt,-7pt){\wlin*\a*(-1,0)[+]\wlin*\b*(0,1)[-]}}\relax
\box0\kern9pt}}

\def\rectangle#1(#2,#3){\relax
\lin#1(#2,0)\lin#1(0,#3)\mov#1(0,#3){\lin#1(#2,0)}\mov#1(#2,0){\lin#1(0,#3)}}

\def\dashrectangle#1(#2,#3){\dashlin#1(#2,0)\dashlin#1(0,#3)\relax
\mov#1(0,#3){\dashlin#1(#2,0)}\mov#1(#2,0){\dashlin#1(0,#3)}}

\def\waverectangle#1(#2,#3){\L*=#1\Lengthunit\a*=#2\L*\b*=#3\L*
\ifdim\a*<0pt\a*-\a*\def\x*{-1}\else\def\x*{1}\fi
\ifdim\b*<0pt\b*-\b*\def\y*{-1}\else\def\y*{1}\fi
\wlin*\a*(\x*,0)[-]\wlin*\b*(0,\y*)[+]\relax
\mov#1(0,#3){\wlin*\a*(\x*,0)[+]}\mov#1(#2,0){\wlin*\b*(0,\y*)[-]}}

\def\calcparab*{\ifnum\n*>\m*\k*\N*\advance\k*-\n*\else\k*\n*\fi
\a*=\the\k* sp\a*=10\a*\b*\dm*\advance\b*-\a*\k*\b*
\a*=\the\*ths\b*\divide\a*\l*\multiply\a*\k*
\divide\a*\l*\k*\*ths\r*\a*\advance\k*-\r*\dt*=\the\k*\L*}

\def\arcto#1(#2,#3)[#4]{\rlap{\toks0={#2}\toks1={#3}\calcnum*#1(#2,#3)\relax
\dm*=135sp\dm*=#1\dm*\d**=#1\Lengthunit\ifdim\dm*<0pt\dm*-\dm*\fi
\multiply\dm*\r*\a*=.3\dm*\a*=#4\a*\ifdim\a*<0pt\a*-\a*\fi
\advance\dm*\a*\N*\dm*\divide\N*10000\relax
\divide\N*2\multiply\N*2\advance\N*\*one
\L*=-.25\d**\L*=#4\L*\divide\d**\N*\divide\L*\*ths
\m*\N*\divide\m*2\dm*=\the\m*5sp\l*\dm*\sm*\n*\*one\loop
\calcparab*\shl**{-\dt*}\advance\n*1\ifnum\n*<\N*\repeat}}

\def\arrarcto#1(#2,#3)[#4]{\L*=#1\Lengthunit\L*=.54\L*
\arcto#1(#2,#3)[#4]\rmov*(#2\L*,#3\L*){\d*=.457\L*\d*=#4\d*\d**-\d*
\rmov*(#3\d**,#2\d*){\arrow.02(#2,#3)}}}

\def\dasharcto#1(#2,#3)[#4]{\rlap{\toks0={#2}\toks1={#3}\relax
\calcnum*#1(#2,#3)\dm*=\the\N*5sp\a*=.3\dm*\a*=#4\a*\ifdim\a*<0pt\a*-\a*\fi
\advance\dm*\a*\N*\dm*
\divide\N*20\multiply\N*2\advance\N*1\d**=#1\Lengthunit
\L*=-.25\d**\L*=#4\L*\divide\d**\N*\divide\L*\*ths
\m*\N*\divide\m*2\dm*=\the\m*5sp\l*\dm*
\sm*\n*\*one\loop\calcparab*
\shl**{-\dt*}\advance\n*1\ifnum\n*>\N*\else\calcparab*
\sh*(#2,#3){\xL*=#3\dt* \yL*=#2\dt*
\rx* \the\cos*\xL* \tmp* \the\sin*\yL* \advance\rx*\tmp*
\ry* \the\cos*\yL* \tmp* \the\sin*\xL* \advance\ry*-\tmp*
\kern\rx*\lower\ry*\hbox{\sm*}}\fi
\advance\n*1\ifnum\n*<\N*\repeat}}

\def\*shl*#1{\c*=\the\n*\d**\advance\c*#1\a**\d*\dt*\advance\d*#1\b**
\a*=\the\toks0\c*\b*=\the\toks1\d*\advance\a*-\b*
\b*=\the\toks1\c*\d*=\the\toks0\d*\advance\b*\d*
\rx* \the\cos*\a* \tmp* \the\sin*\b* \advance\rx*-\tmp*
\ry* \the\cos*\b* \tmp* \the\sin*\a* \advance\ry*\tmp*
\raise\ry*\rlap{\kern\rx*\unhcopy\spl*}}

\def\calcnormal*#1{\b**=10000sp\a**\b**\k*\n*\advance\k*-\m*
\multiply\a**\k*\divide\a**\m*\a**=#1\a**\ifdim\a**<0pt\a**-\a**\fi
\ifdim\a**>\b**\d*=.96\a**\advance\d*.4\b**
\else\d*=.96\b**\advance\d*.4\a**\fi
\d*=.01\d*\r*\d*\divide\a**\r*\divide\b**\r*
\ifnum\k*<0\a**-\a**\fi\d*=#1\d*\ifdim\d*<0pt\b**-\b**\fi
\k*\a**\a**=\the\k*\dd*\k*\b**\b**=\the\k*\dd*}

\def\wavearcto#1(#2,#3)[#4]{\rlap{\toks0={#2}\toks1={#3}\relax
\calcnum*#1(#2,#3)\c*=\the\N*5sp\a*=.4\c*\a*=#4\a*\ifdim\a*<0pt\a*-\a*\fi
\advance\c*\a*\N*\c*\divide\N*20\multiply\N*2\advance\N*-1\multiply\N*4\relax
\d**=#1\Lengthunit\dd*=.012\d**
\divide\dd*\*ths \multiply\dd*\magnitude
\ifdim\d**<0pt\d**-\d**\fi\L*=.25\d**
\divide\d**\N*\divide\dd*\N*\L*=#4\L*\divide\L*\*ths
\m*\N*\divide\m*2\dm*=\the\m*0sp\l*\dm*
\sm*\n*\*one\loop\calcnormal*{#4}\calcparab*
\*shl*{1}\advance\n*\*one\calcparab*
\*shl*{1.3}\advance\n*\*one\calcparab*
\*shl*{1}\advance\n*2\dd*-\dd*\ifnum\n*<\N*\repeat\n*\N*\shl**{0pt}}}

\def\triangarcto#1(#2,#3)[#4]{\rlap{\toks0={#2}\toks1={#3}\relax
\calcnum*#1(#2,#3)\c*=\the\N*5sp\a*=.4\c*\a*=#4\a*\ifdim\a*<0pt\a*-\a*\fi
\advance\c*\a*\N*\c*\divide\N*20\multiply\N*2\advance\N*-1\multiply\N*2\relax
\d**=#1\Lengthunit\dd*=.012\d**
\divide\dd*\*ths \multiply\dd*\magnitude
\ifdim\d**<0pt\d**-\d**\fi\L*=.25\d**
\divide\d**\N*\divide\dd*\N*\L*=#4\L*\divide\L*\*ths
\m*\N*\divide\m*2\dm*=\the\m*0sp\l*\dm*
\sm*\n*\*one\loop\calcnormal*{#4}\calcparab*
\*shl*{1}\advance\n*2\dd*-\dd*\ifnum\n*<\N*\repeat\n*\N*\shl**{0pt}}}

\def\hr*#1{\L*=\xscale\Lengthunit\ifnum
\angle**=0\clap{\vrule width#1\L* height.1pt}\else
\L*=#1\L*\L*=.5\L*\rmov*(-\L*,0pt){\sm*}\rmov*(\L*,0pt){\sl*}\fi}

\def\shade#1[#2]{\rlap{\Lengthunit=#1\Lengthunit
\special{em:linewidth .001pt}\relax
\mov(0,#2.05){\hr*{.994}}\mov(0,#2.1){\hr*{.980}}\relax
\mov(0,#2.15){\hr*{.953}}\mov(0,#2.2){\hr*{.916}}\relax
\mov(0,#2.25){\hr*{.867}}\mov(0,#2.3){\hr*{.798}}\relax
\mov(0,#2.35){\hr*{.715}}\mov(0,#2.4){\hr*{.603}}\relax
\mov(0,#2.45){\hr*{.435}}\special{em:linewidth \the\linwid*}}}

\def\dshade#1[#2]{\rlap{\special{em:linewidth .001pt}\relax
\Lengthunit=#1\Lengthunit\if#2-\def\t*{+}\else\def\t*{-}\fi
\mov(0,\t*.025){\relax
\mov(0,#2.05){\hr*{.995}}\mov(0,#2.1){\hr*{.988}}\relax
\mov(0,#2.15){\hr*{.969}}\mov(0,#2.2){\hr*{.937}}\relax
\mov(0,#2.25){\hr*{.893}}\mov(0,#2.3){\hr*{.836}}\relax
\mov(0,#2.35){\hr*{.760}}\mov(0,#2.4){\hr*{.662}}\relax
\mov(0,#2.45){\hr*{.531}}\mov(0,#2.5){\hr*{.320}}\relax
\special{em:linewidth \the\linwid*}}}}

\def\vdot{\rlap{\kern-1.9pt\lower1.8pt\hbox{$\scriptstyle\bullet$}}}
\def\vtimes{\rlap{\kern-3pt\lower1.8pt\hbox{$\scriptstyle\times$}}}
\def\vDot{\rlap{\kern-2.3pt\lower2.7pt\hbox{$\bullet$}}}
\def\vTimes{\rlap{\kern-3.6pt\lower2.4pt\hbox{$\times$}}}

\def\arc(#1)[#2,#3]{{\k*=#2\l*=#3\m*=\l*
\advance\m*-6\ifnum\k*>\l*\relax\else
{\rotate(#2)\mov(#1,0){\sm*}}\loop
\ifnum\k*<\m*\advance\k*5{\rotate(\k*)\mov(#1,0){\sl*}}\repeat
{\rotate(#3)\mov(#1,0){\sl*}}\fi}}

\def\dasharc(#1)[#2,#3]{{\k**=#2\n*=#3\advance\n*-1\advance\n*-\k**
\L*=1000sp\L*#1\L* \multiply\L*\n* \multiply\L*\Nhalfperiods
\divide\L*57\N*\L* \divide\N*2000\ifnum\N*=0\N*1\fi
\r*\n*  \divide\r*\N* \ifnum\r*<2\r*2\fi
\m**\r* \divide\m**2 \l**\r* \advance\l**-\m** \N*\n* \divide\N*\r*
\k**\r* \multiply\k**\N* \dn*\n* \advance\dn*-\k** \divide\dn*2\advance\dn*\*one
\r*\l** \divide\r*2\advance\dn*\r* \advance\N*-2\k**#2\relax
\ifnum\l**<6{\rotate(#2)\mov(#1,0){\sm*}}\advance\k**\dn*
{\rotate(\k**)\mov(#1,0){\sl*}}\advance\k**\m**
{\rotate(\k**)\mov(#1,0){\sm*}}\loop
\advance\k**\l**{\rotate(\k**)\mov(#1,0){\sl*}}\advance\k**\m**
{\rotate(\k**)\mov(#1,0){\sm*}}\advance\N*-1\ifnum\N*>0\repeat
{\rotate(#3)\mov(#1,0){\sl*}}\else\advance\k**\dn*
\arc(#1)[#2,\k**]\loop\advance\k**\m** \r*\k**
\advance\k**\l** {\arc(#1)[\r*,\k**]}\relax
\advance\N*-1\ifnum\N*>0\repeat
\advance\k**\m**\arc(#1)[\k**,#3]\fi}}

\def\triangarc#1(#2)[#3,#4]{{\k**=#3\n*=#4\advance\n*-\k**
\L*=1000sp\L*#2\L* \multiply\L*\n* \multiply\L*\Nhalfperiods
\divide\L*57\N*\L* \divide\N*1000\ifnum\N*=0\N*1\fi
\d**=#2\Lengthunit \d*\d** \divide\d*57\multiply\d*\n*
\r*\n*  \divide\r*\N* \ifnum\r*<2\r*2\fi
\m**\r* \divide\m**2 \l**\r* \advance\l**-\m** \N*\n* \divide\N*\r*
\dt*\d* \divide\dt*\N* \dt*.5\dt* \dt*#1\dt*
\divide\dt*1000\multiply\dt*\magnitude
\k**\r* \multiply\k**\N* \dn*\n* \advance\dn*-\k** \divide\dn*2\relax
\r*\l** \divide\r*2\advance\dn*\r* \advance\N*-1\k**#3\relax
{\rotate(#3)\mov(#2,0){\sm*}}\advance\k**\dn*
{\rotate(\k**)\mov(#2,0){\sl*}}\advance\k**-\m**\advance\l**\m**\loop\dt*-\dt*
\d*\d** \advance\d*\dt*
\advance\k**\l**{\rotate(\k**)\rmov*(\d*,0pt){\sl*}}%
\advance\N*-1\ifnum\N*>0\repeat\advance\k**\m**
{\rotate(\k**)\mov(#2,0){\sl*}}{\rotate(#4)\mov(#2,0){\sl*}}}}

\def\wavearc#1(#2)[#3,#4]{{\k**=#3\n*=#4\advance\n*-\k**
\L*=4000sp\L*#2\L* \multiply\L*\n* \multiply\L*\Nhalfperiods
\divide\L*57\N*\L* \divide\N*1000\ifnum\N*=0\N*1\fi
\d**=#2\Lengthunit \d*\d** \divide\d*57\multiply\d*\n*
\r*\n*  \divide\r*\N* \ifnum\r*=0\r*1\fi
\m**\r* \divide\m**2 \l**\r* \advance\l**-\m** \N*\n* \divide\N*\r*
\dt*\d* \divide\dt*\N* \dt*.7\dt* \dt*#1\dt*
\divide\dt*1000\multiply\dt*\magnitude
\k**\r* \multiply\k**\N* \dn*\n* \advance\dn*-\k** \divide\dn*2\relax
\divide\N*4\advance\N*-1\k**#3\relax
{\rotate(#3)\mov(#2,0){\sm*}}\advance\k**\dn*
{\rotate(\k**)\mov(#2,0){\sl*}}\advance\k**-\m**\advance\l**\m**\loop\dt*-\dt*
\d*\d** \advance\d*\dt* \dd*\d** \advance\dd*1.3\dt*
\advance\k**\r*{\rotate(\k**)\rmov*(\d*,0pt){\sl*}}\relax
\advance\k**\r*{\rotate(\k**)\rmov*(\dd*,0pt){\sl*}}\relax
\advance\k**\r*{\rotate(\k**)\rmov*(\d*,0pt){\sl*}}\relax
\advance\k**\r*
\advance\N*-1\ifnum\N*>0\repeat\advance\k**\m**
{\rotate(\k**)\mov(#2,0){\sl*}}{\rotate(#4)\mov(#2,0){\sl*}}}}

\def\gmov*#1(#2,#3)#4{\rlap{\L*=#1\Lengthunit
\xL*=#2\L* \yL*=#3\L*
\rx* \gcos*\xL* \tmp* \gsin*\yL* \advance\rx*-\tmp*
\ry* \gcos*\yL* \tmp* \gsin*\xL* \advance\ry*\tmp*
\rx*=\xscale\rx* \ry*=\yscale\ry*
\xL* \the\cos*\rx* \tmp* \the\sin*\ry* \advance\xL*-\tmp*
\yL* \the\cos*\ry* \tmp* \the\sin*\rx* \advance\yL*\tmp*
\kern\xL*\raise\yL*\hbox{#4}}}

\def\rgmov*(#1,#2)#3{\rlap{\xL*#1\yL*#2\relax
\rx* \gcos*\xL* \tmp* \gsin*\yL* \advance\rx*-\tmp*
\ry* \gcos*\yL* \tmp* \gsin*\xL* \advance\ry*\tmp*
\rx*=\xscale\rx* \ry*=\yscale\ry*
\xL* \the\cos*\rx* \tmp* \the\sin*\ry* \advance\xL*-\tmp*
\yL* \the\cos*\ry* \tmp* \the\sin*\rx* \advance\yL*\tmp*
\kern\xL*\raise\yL*\hbox{#3}}}

\def\Earc(#1)[#2,#3][#4,#5]{{\k*=#2\l*=#3\m*=\l*
\advance\m*-6\ifnum\k*>\l*\relax\else\def\xscale{#4}\def\yscale{#5}\relax
{\angle**0\rotate(#2)}\gmov*(#1,0){\sm*}\loop
\ifnum\k*<\m*\advance\k*5\relax
{\angle**0\rotate(\k*)}\gmov*(#1,0){\sl*}\repeat
{\angle**0\rotate(#3)}\gmov*(#1,0){\sl*}\relax
\def\xscale{1}\def\yscale{1}\fi}}

\def\dashEarc(#1)[#2,#3][#4,#5]{{\k**=#2\n*=#3\advance\n*-1\advance\n*-\k**
\L*=1000sp\L*#1\L* \multiply\L*\n* \multiply\L*\Nhalfperiods
\divide\L*57\N*\L* \divide\N*2000\ifnum\N*=0\N*1\fi
\r*\n*  \divide\r*\N* \ifnum\r*<2\r*2\fi
\m**\r* \divide\m**2 \l**\r* \advance\l**-\m** \N*\n* \divide\N*\r*
\k**\r*\multiply\k**\N* \dn*\n* \advance\dn*-\k** \divide\dn*2\advance\dn*\*one
\r*\l** \divide\r*2\advance\dn*\r* \advance\N*-2\k**#2\relax
\ifnum\l**<6\def\xscale{#4}\def\yscale{#5}\relax
{\angle**0\rotate(#2)}\gmov*(#1,0){\sm*}\advance\k**\dn*
{\angle**0\rotate(\k**)}\gmov*(#1,0){\sl*}\advance\k**\m**
{\angle**0\rotate(\k**)}\gmov*(#1,0){\sm*}\loop
\advance\k**\l**{\angle**0\rotate(\k**)}\gmov*(#1,0){\sl*}\advance\k**\m**
{\angle**0\rotate(\k**)}\gmov*(#1,0){\sm*}\advance\N*-1\ifnum\N*>0\repeat
{\angle**0\rotate(#3)}\gmov*(#1,0){\sl*}\def\xscale{1}\def\yscale{1}\else
\advance\k**\dn* \Earc(#1)[#2,\k**][#4,#5]\loop\advance\k**\m** \r*\k**
\advance\k**\l** {\Earc(#1)[\r*,\k**][#4,#5]}\relax
\advance\N*-1\ifnum\N*>0\repeat
\advance\k**\m**\Earc(#1)[\k**,#3][#4,#5]\fi}}

\def\triangEarc#1(#2)[#3,#4][#5,#6]{{\k**=#3\n*=#4\advance\n*-\k**
\L*=1000sp\L*#2\L* \multiply\L*\n* \multiply\L*\Nhalfperiods
\divide\L*57\N*\L* \divide\N*1000\ifnum\N*=0\N*1\fi
\d**=#2\Lengthunit \d*\d** \divide\d*57\multiply\d*\n*
\r*\n*  \divide\r*\N* \ifnum\r*<2\r*2\fi
\m**\r* \divide\m**2 \l**\r* \advance\l**-\m** \N*\n* \divide\N*\r*
\dt*\d* \divide\dt*\N* \dt*.5\dt* \dt*#1\dt*
\divide\dt*1000\multiply\dt*\magnitude
\k**\r* \multiply\k**\N* \dn*\n* \advance\dn*-\k** \divide\dn*2\relax
\r*\l** \divide\r*2\advance\dn*\r* \advance\N*-1\k**#3\relax
\def\xscale{#5}\def\yscale{#6}\relax
{\angle**0\rotate(#3)}\gmov*(#2,0){\sm*}\advance\k**\dn*
{\angle**0\rotate(\k**)}\gmov*(#2,0){\sl*}\advance\k**-\m**
\advance\l**\m**\loop\dt*-\dt* \d*\d** \advance\d*\dt*
\advance\k**\l**{\angle**0\rotate(\k**)}\rgmov*(\d*,0pt){\sl*}\relax
\advance\N*-1\ifnum\N*>0\repeat\advance\k**\m**
{\angle**0\rotate(\k**)}\gmov*(#2,0){\sl*}\relax
{\angle**0\rotate(#4)}\gmov*(#2,0){\sl*}\def\xscale{1}\def\yscale{1}}}

\def\waveEarc#1(#2)[#3,#4][#5,#6]{{\k**=#3\n*=#4\advance\n*-\k**
\L*=4000sp\L*#2\L* \multiply\L*\n* \multiply\L*\Nhalfperiods
\divide\L*57\N*\L* \divide\N*1000\ifnum\N*=0\N*1\fi
\d**=#2\Lengthunit \d*\d** \divide\d*57\multiply\d*\n*
\r*\n*  \divide\r*\N* \ifnum\r*=0\r*1\fi
\m**\r* \divide\m**2 \l**\r* \advance\l**-\m** \N*\n* \divide\N*\r*
\dt*\d* \divide\dt*\N* \dt*.7\dt* \dt*#1\dt*
\divide\dt*1000\multiply\dt*\magnitude
\k**\r* \multiply\k**\N* \dn*\n* \advance\dn*-\k** \divide\dn*2\relax
\divide\N*4\advance\N*-1\k**#3\def\xscale{#5}\def\yscale{#6}\relax
{\angle**0\rotate(#3)}\gmov*(#2,0){\sm*}\advance\k**\dn*
{\angle**0\rotate(\k**)}\gmov*(#2,0){\sl*}\advance\k**-\m**
\advance\l**\m**\loop\dt*-\dt*
\d*\d** \advance\d*\dt* \dd*\d** \advance\dd*1.3\dt*
\advance\k**\r*{\angle**0\rotate(\k**)}\rgmov*(\d*,0pt){\sl*}\relax
\advance\k**\r*{\angle**0\rotate(\k**)}\rgmov*(\dd*,0pt){\sl*}\relax
\advance\k**\r*{\angle**0\rotate(\k**)}\rgmov*(\d*,0pt){\sl*}\relax
\advance\k**\r*
\advance\N*-1\ifnum\N*>0\repeat\advance\k**\m**
{\angle**0\rotate(\k**)}\gmov*(#2,0){\sl*}\relax
{\angle**0\rotate(#4)}\gmov*(#2,0){\sl*}\def\xscale{1}\def\yscale{1}}}

\newcount\CatcodeOfAtSign
\CatcodeOfAtSign=\the\catcode`\@
\catcode`\@=11
\def\@arc#1[#2][#3]{\rlap{\Lengthunit=#1\Lengthunit
\sm*\l*arc(#2.1914,#3.0381)[#2][#3]\relax
\mov(#2.1914,#3.0381){\l*arc(#2.1622,#3.1084)[#2][#3]}\relax
\mov(#2.3536,#3.1465){\l*arc(#2.1084,#3.1622)[#2][#3]}\relax
\mov(#2.4619,#3.3086){\l*arc(#2.0381,#3.1914)[#2][#3]}}}

\def\dash@arc#1[#2][#3]{\rlap{\Lengthunit=#1\Lengthunit
\d*arc(#2.1914,#3.0381)[#2][#3]\relax
\mov(#2.1914,#3.0381){\d*arc(#2.1622,#3.1084)[#2][#3]}\relax
\mov(#2.3536,#3.1465){\d*arc(#2.1084,#3.1622)[#2][#3]}\relax
\mov(#2.4619,#3.3086){\d*arc(#2.0381,#3.1914)[#2][#3]}}}

\def\wave@arc#1[#2][#3]{\rlap{\Lengthunit=#1\Lengthunit
\w*lin(#2.1914,#3.0381)\relax
\mov(#2.1914,#3.0381){\w*lin(#2.1622,#3.1084)}\relax
\mov(#2.3536,#3.1465){\w*lin(#2.1084,#3.1622)}\relax
\mov(#2.4619,#3.3086){\w*lin(#2.0381,#3.1914)}}}

\def\bezier#1(#2,#3)(#4,#5)(#6,#7){\N*#1\l*\N* \advance\l*\*one
\d* #4\Lengthunit \advance\d* -#2\Lengthunit \multiply\d* \*two
\b* #6\Lengthunit \advance\b* -#2\Lengthunit
\advance\b*-\d* \divide\b*\N*
\d** #5\Lengthunit \advance\d** -#3\Lengthunit \multiply\d** \*two
\b** #7\Lengthunit \advance\b** -#3\Lengthunit
\advance\b** -\d** \divide\b**\N*
\mov(#2,#3){\sm*{\loop\ifnum\m*<\l*
\a*\m*\b* \advance\a*\d* \divide\a*\N* \multiply\a*\m*
\a**\m*\b** \advance\a**\d** \divide\a**\N* \multiply\a**\m*
\rmov*(\a*,\a**){\unhcopy\spl*}\advance\m*\*one\repeat}}}

\catcode`\*=12

\newcount\n@ast
\def\n@ast@#1{\n@ast0\relax\get@ast@#1\end}
\def\get@ast@#1{\ifx#1\end\let\next\relax\else
\ifx#1*\advance\n@ast1\fi\let\next\get@ast@\fi\next}

\newif\if@up \newif\if@dwn
\def\up@down@#1{\@upfalse\@dwnfalse
\if#1u\@uptrue\fi\if#1U\@uptrue\fi\if#1+\@uptrue\fi
\if#1d\@dwntrue\fi\if#1D\@dwntrue\fi\if#1-\@dwntrue\fi}

\def\halfcirc#1(#2)[#3]{{\Lengthunit=#2\Lengthunit\up@down@{#3}\relax
\if@up\mov(0,.5){\@arc[-][-]\@arc[+][-]}\fi
\if@dwn\mov(0,-.5){\@arc[-][+]\@arc[+][+]}\fi
\def\lft{\mov(0,.5){\@arc[-][-]}\mov(0,-.5){\@arc[-][+]}}\relax
\def\rght{\mov(0,.5){\@arc[+][-]}\mov(0,-.5){\@arc[+][+]}}\relax
\if#3l\lft\fi\if#3L\lft\fi\if#3r\rght\fi\if#3R\rght\fi
\n@ast@{#1}\relax
\ifnum\n@ast>0\if@up\shade[+]\fi\if@dwn\shade[-]\fi\fi
\ifnum\n@ast>1\if@up\dshade[+]\fi\if@dwn\dshade[-]\fi\fi}}

\def\halfdashcirc(#1)[#2]{{\Lengthunit=#1\Lengthunit\up@down@{#2}\relax
\if@up\mov(0,.5){\dash@arc[-][-]\dash@arc[+][-]}\fi
\if@dwn\mov(0,-.5){\dash@arc[-][+]\dash@arc[+][+]}\fi
\def\lft{\mov(0,.5){\dash@arc[-][-]}\mov(0,-.5){\dash@arc[-][+]}}\relax
\def\rght{\mov(0,.5){\dash@arc[+][-]}\mov(0,-.5){\dash@arc[+][+]}}\relax
\if#2l\lft\fi\if#2L\lft\fi\if#2r\rght\fi\if#2R\rght\fi}}

\def\halfwavecirc(#1)[#2]{{\Lengthunit=#1\Lengthunit\up@down@{#2}\relax
\if@up\mov(0,.5){\wave@arc[-][-]\wave@arc[+][-]}\fi
\if@dwn\mov(0,-.5){\wave@arc[-][+]\wave@arc[+][+]}\fi
\def\lft{\mov(0,.5){\wave@arc[-][-]}\mov(0,-.5){\wave@arc[-][+]}}\relax
\def\rght{\mov(0,.5){\wave@arc[+][-]}\mov(0,-.5){\wave@arc[+][+]}}\relax
\if#2l\lft\fi\if#2L\lft\fi\if#2r\rght\fi\if#2R\rght\fi}}

\catcode`\*=11

\def\Circle#1(#2){\halfcirc#1(#2)[u]\halfcirc#1(#2)[d]\n@ast@{#1}\relax
\ifnum\n@ast>0\L*=\xscale\Lengthunit
\ifnum\angle**=0\clap{\vrule width#2\L* height.1pt}\else
\L*=#2\L*\L*=.5\L*\special{em:linewidth .001pt}\relax
\rmov*(-\L*,0pt){\sm*}\rmov*(\L*,0pt){\sl*}\relax
\special{em:linewidth \the\linwid*}\fi\fi}

\catcode`\*=12

\def\wavecirc(#1){\halfwavecirc(#1)[u]\halfwavecirc(#1)[d]}

\def\dashcirc(#1){\halfdashcirc(#1)[u]\halfdashcirc(#1)[d]}

\def\xscale{1}
\def\yscale{1}

\def\Ellipse#1(#2)[#3,#4]{\def\xscale{#3}\def\yscale{#4}\relax
\Circle#1(#2)\def\xscale{1}\def\yscale{1}}

\def\dashEllipse(#1)[#2,#3]{\def\xscale{#2}\def\yscale{#3}\relax
\dashcirc(#1)\def\xscale{1}\def\yscale{1}}

\def\waveEllipse(#1)[#2,#3]{\def\xscale{#2}\def\yscale{#3}\relax
\wavecirc(#1)\def\xscale{1}\def\yscale{1}}

\def\halfEllipse#1(#2)[#3][#4,#5]{\def\xscale{#4}\def\yscale{#5}\relax
\halfcirc#1(#2)[#3]\def\xscale{1}\def\yscale{1}}

\def\halfdashEllipse(#1)[#2][#3,#4]{\def\xscale{#3}\def\yscale{#4}\relax
\halfdashcirc(#1)[#2]\def\xscale{1}\def\yscale{1}}

\def\halfwaveEllipse(#1)[#2][#3,#4]{\def\xscale{#3}\def\yscale{#4}\relax
\halfwavecirc(#1)[#2]\def\xscale{1}\def\yscale{1}}

\catcode`\@=\the\CatcodeOfAtSign

\maketitle

\begin{abstract}
We review the basic results concerning the structure of
effective action in ${\cal N}=4$ supersymmetric Yang-Mills theory in
Coulomb phase. Various classical formulations
of this theory are considered. We show that the low-energy effective
action depending on all fileds of ${\cal N}=4$ vector multiplet can be
exactly found. This result is discussed on the base of algebraic
analysis exploring the general harmonic superspace techniques and on
the base of straightforward quantum field theory calculations using the
${\cal N}=2$ supersymmetric background field method. We study the
one-loop effective action beyond leading low-energy
approximation and construct supersymmetric generalization
of Heisenberg-Euler-Schwinger effective action depending on all fields
of ${\cal N}=4$ vector multiplet. We also consider the derivation of
leading low-enrgy effective action at two loops.  \
\end{abstract}
\thispagestyle{empty}

\newcommand{\be}{\begin{equation}}
\newcommand{\ee}{\end{equation}}
\newcommand{\bea}{\begin{eqnarray}}
\newcommand{\eea}{\end{eqnarray}}

\section{Introduction}

Four dimensional ${\cal N}=4$ supersymmetric Yang-Mills (SYM) theory
possesses the remarkable properties in quantum domain (see e.g. the
reviews \cite{Soh}, \cite{K}, \cite{HF}):

({\bf {i}}) ${\cal N}=4$ SYM theory is the maximally extended rigid
supersymmetric field model.  It means, if ${\cal N} > 4$, the
supermultiplet includes the fields with spins greater then 1.
Consistent description of their interaction demands to take into
account at least gravity, i.e. to localize the supersymmetry.

({\bf {ii}}) ${\cal N}=4$ SYM theory is the finite four dimensional
quantum field model. The theory under consideration has the only
coupling (gauge coupling $g$) demanding no any infinite
renormalization. The corresponding beta-function vanishes.

({\bf {iii}}) ${\cal N}=4$ SYM theory is the superconformal invariant
field model. The theory under consideration has no any scale. There are
no explicit mass parameters in Lagrangian and no any scale is generated
by radiative corrections. \\ These properties allows to treat ${\cal
N}=4$ SYM theory as the unique quantum field theory model.

Recently there were found unexpected links between ${\cal N}=4$ SYM
theory and \\ string/brane theory. The ${\cal N}=4$ SYM theory is
closely related to $D3$-branes (see e.g. the review \cite{Te}) and there
exists a conjecture that $D3$-brane interactions in static limit can be
completely described in terms of effective action of ${\cal N}=4$ SYM
theory \cite{ChT}, \cite{BPT}. Moreover, according to
AdS/CFT-correspondence, some correlation functions of gauge invariant
operators in ${\cal N}=4$ SYM theory can be found on the base of
low-energy effective action of type IIB superstring compactified on
$AdS_{5}$ x $S_{5}$ manifold and vise versa (see e.g. the reviews
\cite{A}, \cite{HF}). Such properties allow to
consider ${\cal N}=4$ SYM theory as a part of superstring theory.

This paper is devoted to a brief review of problem of low-energy
effective action in ${\cal N}=4$ SYM theory. We begin with descussing
the diverse formulations of the model.  In particular, we consider the
${\cal N}=1$ superfield formulation and ${\cal N}=2$ harmonic
superspace formualtion. Both these formulations are used further for
study of the various aspects of effective action.  We show that the
leading low-energy effective action depending on all fields of ${\cal
N}=4$ vector multipelt can be exactly found. Then we study the one-loop
effective action, also depending on all fields of ${\cal N}=4$ vector
multiplet, beyond leading low-energy order. After that, we discuss the
structure of leading low-energy effective action at two loops.

\section{The various formulations of ${\cal N}=4$ SYM theory}

The ${\cal N}=4$ SYM theory is a dynamical field
model of ${\cal N}=4$ vector multiplet. As known, the on-shell spectrum
of ${\cal N}=4$ vector multiplet consists of one vector $A_{m}$, six
real scalars ${\phi}^{I}$ and four Majorana spinors ${\lambda}^{A}$
(see e.g.  \cite{Soh}).  At present, three formulations of
${\cal N}=4$ SYM theory are usually used: component
formulation, formulation in terms of ${\cal N}=1$ superfields and
formulation in terms of ${\cal N}=2$ harmonic superfields.

{\bf Component formulation}. Action of ${\cal N}=4$ SYM theory in terms
of component fields $A_{m}, {\phi}^{I}, {\lambda}^{A}$ has been
constructed in \cite{BSS} by means of dimensional reduction from ten
dimensional ${\cal N}=1$ SYM theory to four dimensions (see also
\cite{GSW} and the reviews \cite {Soh}, \cite{K}, \cite{HF}).
This action is written as follows
\begin{eqnarray}
S &=& \int d^{4}x{\rm tr}
\{-{1\over 2g^{2}}F^{mn}F_{mn} +
{1\over 2} {\cal D}_{m}{\phi}^{I}{\cal D}^{m}{\phi}^{I} -
i\bar{\lambda}^{A} {\sigma}^{m}{\cal D}_{m}{\lambda}_{A} + \nonumber\\
&+& gC^{AB}_{I}{\lambda}_{A}[{\phi}^{I},{\lambda}_{B}] +
g{\bar C}_{IAB}\bar{\lambda}^{A}[{\phi}^{I},\bar{\lambda}^{B}] +
{1\over 2} g^{2}[{\phi}^{I},{\phi}^{J}][{\phi}^{I},{\phi}^{J}]
\}.\label{1}
\end{eqnarray}
Here $g$ is gauge coupling, $C^{AB}_{I}, {\bar C}_{IAB}$ are six-
dimensional analogs of four-dimensional sigma-matrices, the scalar and
spinor fields belong to adgoint representation of the gauge group. One
can show that action (\ref{1}) is invariant under four hidden on-shell
supersymmetries \cite{Soh}, \cite{K}, \cite{HF}.

To describe a structure of graund state of the theory ones consider the
conditions of vanishing the scalar potential in action (\ref{1}). It
has the form ${\rm tr}[{\phi}^{I},{\phi}^{J}][{\phi}^{I},{\phi}^{J}] =
0$. Here ${\phi}^{I} = {\phi}^{I}_{c}T^{c}$ where $T^{c}$ are
the generators of the gauge algebra in adjoint representation. Hence,
the scalar potential vanishes for the fields ${\phi}^{I}_{c}$
satisfying the condition
\begin{equation}
{\phi}^{I}_{c_{1}}{\phi}^{J}_{c_{2}}{\phi}^{I}_{d_{1}}{\phi}^{J}_{d_{2}}{\rm
tr}([T^{c_{1}},T^{c_{2}}][T^{d_{1}},T^{d_{2}}]) = 0. \label{2}
\end{equation}
Let $r$ is a rank of gauge group. Then for each fixed index $I$,
the $r$ components ${\phi}^{I}_{c}$ can have nonzero values. If all of
them are nonzero, the gauge group is broken down to its
maximal Abelian subgroup. In particular, if the gauge group is
$SU(N)$, we get group $U(1)^{N-1}$ after symmetry breaking. This case
is called Coulomb phase of the theory under consderation. Further we
study the ${\cal N}=4$ SYM theory just in Coulomb phase.

{\bf Formulation in terms of ${\cal N}=1$ superfiedls}. On-shell
${\cal N}=4$ vector multiplet can be decomposed into a sum of
on-shell ${\cal N}=1$ vector multiplet and three scalar multiplets.
 Each of these ${\cal N}=1$ multiplets is
formulated in superfield terms on ${\cal N}=1$ superspace. It means,
there exists a possibility to formulate ${\cal N}=4$ SYM theory in
terms of real scalar ${\cal N}=1$ superfield $V$ and three chiral
$\Phi^{i}$ and antichiral $\bar\Phi_{i}$ superfields.  The
corresponding action is written as follows \cite{GGRS}
\begin{eqnarray}
S &=& {1\over g^{2}}{\rm tr}\{\int
d^{4}x d^{2}\theta\,W^{2}+
\int d^{4}x d^{4}\theta\,\bar{\Phi}_{i}{\rm
e}^{V}\Phi^{i}{\rm e}^{-V}+\nonumber\\
&+& {1\over 3!}\int d^{4}x
d^{2}\theta\, i{\epsilon}_{ijk}\Phi^{i}[\Phi^{j},\Phi^{k}]+{1\over
3!}\int d^{4}x d^{2}\bar\theta\,
i{\epsilon}^{ijk}\bar\Phi_{i}[\bar\Phi_{j},\bar\Phi_{k}] \}.\label{3}
\end{eqnarray}
Here $W_{\alpha}$ is ${\cal
N}=1$ superfield strength, all superfields are taken in the adjoint
representation of the gauge group. The action (\ref{3}) possesses the
manifest ${\cal N}=1$ supersymmetry. However this action has
three on-shell hidden global supersymmetries (\cite{GGRS}, see also
\cite{BBP1}) and as a result ones get ${\cal N}=4$
supersymmetric theory.

{\bf Formulation in terms of ${\cal N}=2$ harmonic superspace}. Another
way of considering the ${\cal N}=4$ vector multiplet is to
decompose it into a sum of ${\cal N}=2$ vector multiplet and
hypermultiplet.  Both these ${\cal N}=2$ multiplets can be formulated
in terms of superfields defined on harmonic superspace
\cite{GIKOS}, \cite{GIOS}. In harmonic superspace
approach, the ${\cal N}=2$ vector multiplet is described by analytic
superfield $V^{++}$ and the hypermultiplet is described by analytic
superfield $q^{+}$. It means, there can exist a
formulation of the ${\cal N}=4$ SYM theory in terms of the superfields
$V^{++}, q^{+}$. Such a formulation actually exists. The
corresponding action is written as follows \cite{GIOS}
\begin{eqnarray}
S &=& {1\over {2g^2}}{\rm tr}\int d^{8}z{\cal W}^2 - {1\over {2}}{\rm
tr}\int d{\zeta}^{(-4)} q^{+a}(D^{++}+iV^{++})q^{+}_{a}. \label{4}
\end{eqnarray}
Here $d^{8}z$ is integration measure over chiral subspace of
general ${\cal N}=2$ superspace, $d{\zeta}^{(-4)}$ is integration
measure over analytic subspace of harmonic superspace, $q_{a} =
(q^{+},{\tilde {q}}^{+})$, $q^{a} = \epsilon^{ab}q_{b}$ is the
hypermultiplet analytic superfield in the adjoint representation of the
gauge group and ${\cal W}$ is the ${\cal N}=2$ superfield strength of
analytic superfield $V^{++}$. The action (\ref{4}) is manifestly ${\cal
N}=2$ supersymmetric. However it possesses the hidden on-shell
${\cal N}=2$ supersymmetry and as a result ones get ${\cal
N}=4$ supersymmetric theory.

\section{Exact low-energy effective action.}

Possibility of constructing the exact low-energy effective action in
${\cal N}=4$ SYM theory is direct consequence of the strong restrictions
imposed by maximally extended rigid supersymmetry, $R$-symmetry and
scale invariance on a structure of effective action. For exploration of
these restrictions it is convenient to use the formulation of the
${\cal N}=4$ SYM theory in terms of harmonic superspace, where the most
number of supersymmetries is manifest in compare with the component and
${\cal N}=1$ superspace formulations. Of course, the same results, in
principle, can be obtained using any of the formluations. The feature
of harmonic superspace formulation is manifest ${\cal N}=2$
supersymmetry which simplify a general consideration.

In harmonic superspace approach, the ${\cal N}=4$ SYM effective action
${\Gamma}$ depends on ${\cal N}=2$ vector multiplet superfield
$V^{++}$ and hypermultiplet superfield $q^{+}$. Due to gauge
invariance, the superfield $V^{++}$ enters into effective action through
the ${\cal N}=2$ superfield strengths ${\cal W}$ and $\bar{\cal W}$.
\footnote{We imply that the quantum theory is formulated within
background field method leading to gauge invariant effective action
(see formulation of background field method in ${\cal N}=2$ harmonic
superspace in \cite{BBKO}, \cite{BKO}).} The effective
action is ${\Gamma}[{\cal W},\bar{\cal W},q^{+}]$ has the
following general form
\begin{eqnarray}
{\Gamma}[{\cal W},\bar{\cal W},q^{+}] &=& S[V^{++},q^{+}] +
\bar{\Gamma}[{\cal W},\bar{\cal W},q^{+}].\label{5}
\end{eqnarray}
where $S[V^{++},q^{+}]$ is the classical action (\ref{4}) and
$\bar{\Gamma}[{\cal W},\bar{\cal W},q^{+}]$ incorporates all quantum
corrections. In low-energy approximation we neglect all spinor and
space-time derivatives what allows to write the $\bar{\Gamma}[{\cal W},
\bar{\cal W},q^{+}]$ in terms of effective Lagrangian ${\cal
L}_{eff}({\cal W},\bar{\cal W},q^{+})$ as follows
\begin{eqnarray}
\bar{\Gamma}[{\cal W},\bar{\cal W},q^{+}] &=& \int d^{12}zdu {\cal
L}_{eff}({\cal W},\bar{\cal W},q^{+}).\label{6}
\end{eqnarray}
Here $d^{12}z$ is the full ${\cal N}=2$ superspace measure and $du$
means integration measure over harmonics. We denote
\begin{eqnarray}
{\cal H}({\cal W},\bar{\cal W}) &=& {\cal L}_{eff}({\cal W},\bar{\cal
W},q^{+})|_{q^{+}=0}.\label{7}
\end{eqnarray}
The quantity ${\cal H}({\cal W},\bar{\cal W})$ determines the
${\cal N}=4$ SYM low-energy effective action in sector of ${\cal
N}=2$ vector multiplet and is called the nonholomorphic
effective potential.  As a result the effective Lagrangian can be
presented in the form
\begin{eqnarray}
{\cal L}_{eff}({\cal
W},\bar{\cal W},q^{+}) &=& {\cal H}({\cal W},\bar{\cal W}) + {\cal
L}_{q}({\cal W},\bar{\cal W},q^{+}).\label{8}
\end{eqnarray}
where the
quantity ${\cal L}_{q}({\cal W},\bar{\cal W},q^{+})$ vanishes at
$q^{+}=0$ and determines the ${\cal N}=4$ SYM low-energy effective
action in the hypermultiplet sector. Further we show that the effective
Lagrangian ${\cal L}_{eff}$ can be exactly found in Coulomb phase where
${\cal W}$ and $q^{+}$ belong to maximal Abelian subroup of the gauge
group.

We begin with $SU(2)$ gauge theory spontaneously broken down to $U(1)$.
First of all, ones consider the calculation of nonholomorphic
effective potential following the work \cite{DS}. Taking into account
the mass dimensions of $d^{12}z$, ${\cal W}$ and $\bar{\cal W}$ ones
see that nonholomorphic effective potential is dimensionless, and hence
\begin{equation}
{\cal H}({\cal W},\bar{\cal W}) = {\cal H}({{\cal
W}\over {\Lambda}}, {\bar{\cal W}\over {\Lambda}}).\label{9}
\end{equation}
with ${\Lambda}$ be some scale. Now we use the conditions
of $R$-invariance and scale independence of ${\cal N}=4$ SYM quantum
theory.  It leads to equation
\begin{equation}
{\Lambda}{d\over
{d{\Lambda}}} \int d^{12}z {\cal H}({{\cal W}\over
{\Lambda}},{\bar{\cal W}\over {\Lambda}} ) = 0.\label{10}
\end{equation}
As shown in \cite{DS}, the
equation (\ref{10}) has the only solution
\begin{eqnarray}
{\cal H}({{\cal W}\over {\Lambda}}, {\bar{\cal W}\over {\Lambda}}) &=& {\rm
c}{\rm ln}{{\cal W}\over {\Lambda}}{\rm ln}{\bar{\cal W}\over
{\Lambda}},\label{11}
\end{eqnarray}
where $c$ is an arbitrary dimensionless constant which can depend only
on gauge coupling $g$. Result (\ref{11}) is exact. Thus, we see that all
quantum corrections, perturbative or non-perturbative, are included
into a single constant $c$. Moreover, it was argued in the work
\cite{DS} (see also \cite{LU}) that the constant $c$ really
gets only one-loop contribution.\footnote{The straightforward
calculations shown that three- and four-loop contributions to $c$ are
absent \cite{BP} in agreement with general statement \cite{DS}.} The
direct calculations of the constant $c$ were carried out in the papers
\cite{H}, \cite{BK}, \cite{BBK}, \cite{LU} (see also the review
\cite{BBIKO} and calculations of the constant $c$ in ${\cal N}=2$
superconformal theories in \cite {WGR}, \cite{KM1}).  Thus, the exact
low-energy effective action in the ${\cal N}=2$ vector multiplet sector
is known.  However, to get complete effective Lagrangian ${\cal
L}_{eff}({\cal W},\bar{\cal W},q^{+})$ we have to find the
hypermultiplet dependent part ${\cal L}_{q}({\cal W},\bar{\cal
W},q^{+})$ in (\ref{8}).

Problem of complete ${\cal L}_{eff}$ has been solved in works
\cite{BI}, \cite{BIP}. We consider the calculation of ${\cal L}_{q}$
following the paper \cite{BI}. First of all, we note that the effective
action (\ref{7}) is manifestly ${\cal N}=2$ supersymmetric. However,
the classical action (\ref{4}) is also invariant under the hidden ${\cal
N}=2$ supersymmetry, forming together with manifest supersymmetry, the
complete ${\cal N}=4$ supersymmetry of ${\cal N}=4$ SYM theory. We will
seek for the function ${\cal L}_{q}$ demanding that the full effective
action (\ref{7}) is invariant under the hidden ${\cal N}=2$
supersymmetry transformations. One can show, it is sufficient to
consider the Abelian background superfields ${\cal W}$, $\bar{\cal W}$,
$q^{+}$ satisfying the free classical equations of motion \cite{BI}.
In this case, the hidden ${\cal N}=2$ supersymmetry transformations have
the form \cite{GIOS}
\begin{eqnarray}
\delta{\cal W} = {1\over {2}}{\bar{\epsilon}}^{{\dot {\alpha}}a}
{\bar{D}}^{-}_{\dot {\alpha}} q^{+}_{a},
\delta{\bar{\cal W}} = {1\over
{2}}{\epsilon}^{{\alpha}a}D^{-}_{{\alpha}} q^{+}_{a},
{\delta}q^{+}_{a} = {1\over
{4}}({\epsilon}^{\beta}_{a}D^{+}_{{\beta}}{\cal W} +
{\bar{\epsilon}}^{\dot{\alpha}}_{a}{\bar{D}}^{+}_{\dot{\alpha}}{\bar{\cal
W}}).\label{12}
\end{eqnarray}
We demand
\begin{equation}
\delta \int d^{12}zdu ({\cal H}({\cal W},\bar{\cal W}) + {\cal
L}_{q}({\cal W},\bar{\cal W},q^{+})) = 0.\label{13}
\end{equation}
where the transformation ${\delta}$ in (\ref{13}) is generated by
(\ref{12}) and nonholomorphic effective potential ${\cal H}({\cal
W},\bar{\cal W})$ is given by (\ref{9}), (\ref{11}). The equation
(\ref{13}) has been analysed in \cite{BI}. The function ${\cal L}_{q}$
was written as a power series in quantity ${q^{+a}q^{-}_{a}\over
 {{\bar{\cal W}}{\cal W}}}$
with unknown coefficients which were found from equation (\ref{13}). It
was proved that there exists the only function ${\cal L}_{q}({\cal
W},\bar{\cal W},q^{+})$ forming, together with given function ${\cal
H}({\cal W},\bar{\cal W})$ (\ref{9}), (\ref{11}), a solution to the
equation (\ref{13}). Final result has the form
\begin{eqnarray}
{\cal L}_{eff} &=& {\rm c}({\rm ln}{{\cal W}\over {\Lambda}}{\rm
ln}{{\bar{\cal W}}\over {\Lambda}} + (X - 1){{\rm ln}(X - 1)\over {X}}
+ ({\rm Li}_{2}(X) - 1)).\label{14}
\end{eqnarray}
where ${\rm Li}_{2}(X)$ is Euler dilogarithm function, $c$ is the same
consatnt as in (\ref{11}) and
\begin{equation}
X = -{q^{ia}q_{ia}\over
{{\bar{\cal W}}{\cal W}}}.\label{15}
\end{equation}
Here on-shell ${\cal N}=2$ superfield $q^{ia}$ is defined by
$q^{+a}=q^{ia}u^{+}_{i}$  \cite{GIOS}.  It means that the quantity $X$
(\ref{15}) is harmonic independent.  Therefore ${\cal L}_{eff}$ does
not depend on harmonics and the integral over harmonics in (\ref{6})
can be omitted.

We emphasize that effective Lagrangian (\ref{14}) determines the
exact ${\cal N}=4$ supersymmetruc low-energy effective action
depending on all fields of ${\cal N}=4$ vector multiplet.
Nonholomorphic effective potential ({\ref{9}), (\ref{11}) is exact
\cite{DS}. The function ${\cal L}_{q}$ was uniquely found on the base
of this potential using invariance of effective action under the hidden
${\cal N}=2$ supersymmetry transformations. Therefore the effective
Lagrangian (\ref{14}) determines the exact low-energy effective action
of ${\cal N}=4$ SYM theory with the $SU(2)$ gauge group spontaneosly
broken down to $U(1)$.

We investigate now a component structure of low-energy effective action
in bosonic sector. Let the on-shell superfield ${\cal W}$ has the
following bosonic components:complex scalar ${\phi}$ and self-dual
spinor component of Abelian strenght $F^{{\alpha}{\beta}}$ and the
on-shell hypermultiplet superfield $q^{ia}$ has the scalar component
$f^{ia}$. Substituting the effective Lagrangian (\ref{14}) depending
on above superfields into effective action (\ref{6}) and integrating
over anticommuting coordinates ones get (see the details in \cite{BI})
\begin{eqnarray}
{\bar{\Gamma}} &=& 4{\rm c} \int d^{4}x {F^{2}{\bar{F}}^{2}\over
{(|{\phi}|^{2} + f^{ia}f_{ia})^{2}}}.\label{16}
\end{eqnarray}
Expression in the denominator is the invarint quadratic combination of
six scalars from ${\cal N}=4$ vector multiplet under $R$-symmetry
group of ${\cal N}=4$ supersymmetry. We could expect such a result
from the very beginning. It means, in particular, that starting with
bosonic component effective action in sector of ${\cal N}=2$ vector
multiplet and demanding the invariance under $R$-symmetry group we
could restore uniquely the complete result (\ref{16}).

The exact low-energy effective Lagrangian (\ref{14}) has been obtained
for $SU(2)$ gauge group spontaneously broken down to $U(1)$.
Generalization for the $SU(N)$ group spontaneously broken down to
$U(1)^{N-1}$ can be done simply enough using the procedure described in
\cite{BBK}. The final result is presented as a sum over roots of
gauge algebra of the terms having the form (\ref{14}) (see the
details in \cite{BI}). No any other contributions to low-energy
effective action in ${\cal N}=4$ SYM theory are possible.\footnote{In
principle, if we take into account only manifest ${\cal N}=2$
supersymmetry and structure of $SU(N)$ group with $N>2$, then there can
be the terms in low-energy effective action different from (\ref{14})
\cite{DG}, \cite{LU}. However they are forbidden by hidden ${\cal
N}=2$ supersymmetry \cite{BI}.} It leads to important conclusion that
we get the exact low-energy effective action in theory with $SU(N)$
gauge group spontaneously broken down to $U(1)^{N-1}$ group for any
$N$.

\section{Supergraph calculation of the effective Lagrangian}
We have constructed the exact ${\cal N}=4$ supersymmetric
low-energy effective action from purely algebraic consideration. One
can show that this result can be obtained in framework of quantum field
theory by calculating the one-loop harmonic supergraphs with an
arbitrary number of the external hypermultiplet legs on the background
of constant superfields ${\cal W}$, $\bar{\cal W}$ \cite{BIP}. We
discuss now the basic elements of this calculation.

The ${\cal N}=4$ SYM theory is formulated in ${\cal N}=2$
harmonic superspace in terms of ${\cal N}=2$ gauge superfield and
hypermultiplet superfield. We consider the theory with gauge group
$SU(2)$ spontaneously broken down to $U(1)$. To provide the gauge
invariance of the quantum theory we use the
manifestly ${\cal N}=2$ supersymmetric background field methods
\cite{BBKO}. In this case the superfields $V^{++}, q^{+}$ are
splitted into background and quantum parts. All quantum superfield
propagators depend on Abelian background superfields ${\cal W},
\bar{\cal W}$ and, as a result, the effective action is gauge invariant
and manifestly ${\cal N}=2$ supersymmetric functional of the ${\cal W},
\bar{\cal W}$ and background hypermultiplet superfields.

Efective Lagrangian ${\cal L}_{eff}$ ({\ref{8}) is determined by
nonholomorphic effective potential \\
${\cal H}({\cal W}, \bar{\cal W})$
and hypermultiplet dependent effective Lagrangian ${\cal L}_{q}$.
Nonholomorphic effective potential has been studied in enough details
(see the review \cite{BBIKO}). We consider here the harmonic supergraph
calculation of the ${\cal L}_{q}$.

The hypermultiplet dependent contributions to the one-loop
effective action are presented by following infinite sequence of the
supergraphs:
\vspace*{5mm}

\hspace*{.5cm}
\Lengthunit=0.95cm
\GRAPH(hsize=3){\mov(0,-.5){
\halfcirc(1.0)[u]\halfwavecirc(1.0)[d]
\mov(-.6,0){\lin(-.5,0)}\mov(.45,0){\lin(.5,0)}}
\ind(18,0){+}\ind(12,-3){2}\ind(-11,-3){1}
}
\GRAPH(hsize=3){\lin(-.7,.7)\wavelin(1.1,0)\lin(0,-1)
\mov(1,-1){\lin(.7,-.7)\lin(0,1)\wavelin(-1.1,0)}\ind(19,-5){+}
\mov(.9,0){\lin(.7,.7)}\mov(-.2,-1){\lin(-.7,-.7)}
\ind(16,5){3}\ind(16,-15){4}\ind(-12,5){2}\ind(-12,-15){1}
}
\hspace*{.5cm}
\GRAPH(hsize=3){\lin(-.7,.7)\wavelin(.6,0)\lin(0,-1)
\mov(.5,0){\lin(0,1)\lin(.45,0)}
\mov(1,-1){\lin(.7,-.7)\wavelin(0,1)\lin(-.5,0)\ind(13,4){+}
\mov(-.6,0){\lin(0,-1)\wavelin(-.6,0)}}
\mov(.9,0){\lin(.7,.7)}\mov(-.2,-1){\lin(-.7,-.7)}
\ind(16,5){4}\ind(16,-15){5}\ind(-12,5){2}\ind(-12,-15){1}
\ind(4,8){3}\ind(4,-18){6}
}
\hspace*{0.5cm}
\GRAPH(hsize=3){\mov(.1,0){\lin(-.7,.7)\wavelin(.55,0)}
\lin(0,-.5)\mov(0,-.5){\wavelin(0,-.5)\lin(-.7,0)}
\mov(.5,0){\lin(0,1)\lin(.5,0)}
\mov(1,-1){\lin(.7,-.7)\lin(0,.5)\wavelin(-.5,0)
\mov(-.6,0){\lin(0,-1)\lin(-.6,0)}}
\mov(.9,-.5){\wavelin(0,.5)\lin(.7,0)}
\mov(.8,0){\lin(.7,.7)}\mov(-.3,-1){\lin(-.7,-.7)}
\ind(16,5){4}\ind(16,-15){8}\ind(-12,5){2}\ind(-12,-15){8}
\ind(4,8){3}\ind(4,-18){7}\ind(17,-5){5}\ind(-15,-5){1}
\ind(21,-5){+}\ind(27,-5){\ldots}
}
\vspace*{4mm}

\noindent
Here the wavy line stands for ${\cal N}=2$ gauge superfield propagator
and solid external and internal lines stand for external background
hypermultiplet superfields and quantum hypermultiplet propagators
respectively. Explicit expressions for the background dependent
propagators are given in \cite{BBKO}, \cite{BIP}. In the case under
consideration it is sufficient to consider the constant background
strengths ${\cal W}, \bar{\cal W}$. Then any of the above supergraphs
can be exactly calculated. The supergraph with $2n, (n=1,2...)$
external hypermultiplet legs has the form
\begin{eqnarray}
\Gamma_{(2n)} &=& {1\over {n^{2}(n+1)}}{1\over {(4\pi)^{2}}}
\int {\rm d}^{12}z X^{n}, \label{X}
\end{eqnarray}
where $X$ is given by (\ref{15}). To find the complete
dependence of effective action on hypermultipelts, we have to sum all
$\Gamma_{(2n)}$ (\ref{X}). It leads to ${\cal L}_{q} =
{\rm c}((X-1)\frac{{\rm ln}(X-1)}{X} + ({\rm Li}_{2}(X)-1))$ with ${\rm
c} = {1\over {(4\pi)^{2}}}$. This result exactly corresponds to
(\ref{14}).

\section{One-loop effective action beyond leading \\low-energy
approximation}
Eqs (\ref{14}, (\ref{15}) determine the effective action in leading
low-energy order. In this section we consider a structure of the
one-loop effective action beyond leading approximation \cite{BBP1}
using formulation (\ref{3}) of ${\cal N}=4$ SYM theory in terms of
${\cal N}=1$ superfields. Although such a formulation
preserves lesser number of manifest supersymmetries then harmonic
superspace approach, it has some positive features of technical
character. In particular, the one-loop effective action in ${\cal N}=1$
superfield theories can be studied by well developed operator methods
(see e.g.  \cite{book}).\footnote{Proper-time method
allowing to find the one-loop effective action in ${\cal
N}=2$ harmonic supespace beyond leading low-energy approximation has
recently been formulated in \cite{KM1}.}

As well as in previous section, we consider the theory with gauge group
$SU(2)$ spontaneously broken down to $U(1)$. Generalization for $SU(N)$
group spontaneously broken down to $U(1)^{N-1}$ can be done using
procedure described in \cite{BBK}. To provide gauge invariance of
effective action we use ${\cal N}=1$ background field method
\cite{book}, \cite{GGRS} and ${\cal N}=1$ supersymmetric
$R_{\xi}$-gauges \cite{OW}.

Background field method begins with splitting the fields into
background and qauntum. First of all ones rename the fields in action
(\ref{3}) as follows: ${\Phi}^{1} = {\Phi}$, ${\Phi}^{2} = Q$,
${\Phi}^{3} = {\tilde Q}$. The ${\cal N}=1$ chiral superfield ${\Phi}$
contains two real scalars corresponding to ${\cal N}=2$
vector multiplet and the ${\cal N}=1$ chiral superfields $Q, \tilde{Q}$
contain four real scalars corresponding to hypermultiplet. After
background-quantum spliting ones get a theory of quantum superfields
$v, {\phi}, q, {\tilde q}$ in the backgroud fields $V, {\Phi}, Q,
{\tilde Q}$.  In one-loop approximation the effective action is
determined on the base of quadratic part (in quantum fields) of
classical action with gauge fixing term and ghost action. It allows to
obtain the one-loop quantum correction to classical action in the form
(see the details in \cite{BBP1})
\begin{eqnarray}
\bar{\Gamma} = i {\rm Tr\,ln}(\Box - iW^{\alpha}\nabla_{\alpha}
-i{\bar{W}}^{\dot{\alpha}}{\bar{\nabla}}_{\dot{\alpha}} -M).\label{17}
\\ M = \bar{\Phi}\Phi+\bar{Q}Q+\bar{\tilde{Q}}\tilde{Q}.  \label{18}
\end{eqnarray}
Here $W^{\alpha}, {\bar{W}}^{\dot{\alpha}}$ are the background
strengths. ${\rm Tr}$ means the functional trace of the operator acting
in space of quantum superfieds $v$.

Effective action (\ref{17}) has been calculated for the ${\cal
N}=1$ supersymmetric background corresponding to constant Abelian
strength $F_{mn}$ and constant scalar field ${\phi}$ in \cite{BKT}
(see also \cite{PB}) using superfield proper-time techniques. This
calculation corresponds to the case when the superfileds $Q,\tilde{Q}$
vanish.  However, since these superfileds enter in effective action in
special $R$-symmetry invariant combination $M$ (\ref{18}), the
effective action for background with constant component fields $F_{mn},
{\phi}, f^{ia}$ is obtained from result given in \cite{BKT}, \cite{PB}
replacing the expression $\bar{\Phi}{\Phi}$ in \cite{BKT}, \cite{PB}
by $M$ (\ref{18}).  The final result for the effective action
(\ref{17}) is written as follows
\begin{eqnarray}
\bar{\Gamma} =
\frac{1}{8\pi^{2}}\int{\rm d}^8z \int_{0}^{\infty}{\rm d}t\,{\rm
e}^{-t}\frac{ W^{2}\bar{W}^{2}}{M^2}\,\omega
(t\Psi,t\bar{\Psi}).\label{19}
\end{eqnarray}
where
\begin{equation}
{\bar \Psi}^2 = \frac{1}{M^2} \,D^2 {W^2}, \quad \Psi^2 =
\frac{1}{M^2} \,{\bar D}^2 {\bar W}^2.\label{20}
\end{equation}
The function $\omega(x,y)$ has been introduced in \cite{BKT} and looks
like
\begin{eqnarray}
\omega(x,y) &=& \frac{{\rm
cosh}(x)-1}{x^{2}}\frac{{\rm cosh}(y)-1}{y^{2}} \frac{x^{2} -
y^{2}}{{\rm cosh}(x) - {\rm cosh}(y)}.\label{21}
\end{eqnarray}
Eq (\ref{19}) determines the one-loop ${\cal N}=4$ SYM effective action
in ${\cal N}=1$ superfield form for constant ${\cal N}=1$
supersymmetric background.

The effective action (\ref{19}) includes dependence on arbitrary powers
of Abelian strength $F_{mn}$, not only on fourth power like in
(\ref{14}), and, hence, determines the one-loop effective action beyond
leading low-energy approximation. It is given by integral over
proper-time $t$, contains background superfields corresponding to
all fields of ${\cal N}=4$ vector multiplet, both bosinic and fermionic.
Therefore this effective action can be considered as a
generalization of Heisenberg-Euler-Schwinger effective action for
${\cal N}=4$ SYM theory.
In component form for the constant bosonic background $F_{mn}$, the
one-loop effective action of ${\cal N}=4$ SYM theory was obtained in
\cite{FT}. Expanding the function $\omega(x,y)$ (\ref{21}) in
power series in $x,y$ we get the expansion of effective action
(\ref{19}) over $\frac{F^2}{M^2}$. It is interesting to note, this
expansion does not include the $F^6$ term, that is a feature just of
${\cal N}=4$ SYM theory \cite{FT}, \cite{BKT}. However, such a
term appears in expansion of effective action of arbitrary ${\cal
N}=2$ superconformal model \cite{BKT}.

The effective action (\ref{19}) is given in manyfestly ${\cal N}=1$
supersymmetric form. It was shown in \cite{BBP1} that each term in
expansion of this effective action (\ref{19}) over $\frac{F^2}{M^2}$
can be written in manifestly ${\cal N}=2$ sypersymmetric form. To do
that ones introduce the ${\cal N}=2$ superconformal scalars \cite{BKT}
\begin{equation}
{\bf {\bar{\Psi}}} = {1\over {\bar{\cal W}}^{2}} D^{4}{\rm
ln}{\cal W}, \quad {\bf {\Psi}} = {1\over {\cal W}^{2}}
{\bar{D}}^{4}{\rm ln}{\bar{\cal W}}.\label{22}
\end{equation}
Using the expansion of function $\omega(x,y)$ (\ref{21}),
the ${\cal N}=1$ superfields (\ref{20}) and applying the procedure of
restorating the ${\cal N}=2$ strength superfields ${\cal W}, \bar{\cal
W}$ ones get manifestly ${\cal N}=2$ supersymmetric expansion of
effective action (\ref{19}) in power series in the quantities (\ref{22})
\begin{equation}
\bar{\Gamma} = \Gamma^{(0)} + \Gamma^{(2)}) + \Gamma^{(3)} + \cdots.
\label{23}
\end{equation}
Here the
$\Gamma^{(n)}$ contains the terms ${\bf
{\Psi}}^{m}$${\bf {\bar{\Psi}}}^{n-m}$ what corresponds to
$\frac{F^{4+2n}}{M^{2+2n}}$ in bosonic component sector. A few first
terms in the expansion (\ref{23}) can be easy found.  Leading
low-energy contribution is
\begin{eqnarray}
\Gamma^{(0)} = {1\over
{(4\pi)^2}} \int {\rm d}^{12}z \,\left({\rm ln}\frac{{\cal
W}}{{\Lambda}} {\rm ln}\frac{\bar{\cal W}}{{\Lambda}} +
\sum_{k=1}^{\infty}{1\over {k^2(k+1)}}X^{k}\right),\label{24}
\end{eqnarray}
where $X$ is given by (\ref{15}). The series in right hand side of
({\ref{24}) can be sumed up and the result is the hypermultiplet
dependent effective Lagrangian ${\cal L}_{q} = {1\over
{(4\pi)^{2}}}((X-1)\frac{{\rm ln}(X-1)}{X} + ({\rm Li}_{2}(X)-1))$ in
(\ref{8}).  Thus, the $\Gamma^{(0)}$ (\ref{24}) coincides with exact
low-energy effective action (\ref{6}), (\ref{14}).  The terms
$\Gamma^{(2)}, \Gamma^{(3)}, \cdots$ present the subleading corrections
to leading contribution (\ref{24}). The explicit expressions for
$\Gamma^{(2)}, \Gamma^{(3)}, \Gamma^{(4)}$ were calculated in
\cite{BBP1} (see also \cite{BBP2}).

\section{Low-energy effective action at two loops}
In this section we consider a structure of leading two-loop
contribution to effective action. As we discussed in Section 3, the
exact low-energy effective action contains $F^4$ term in bosonic
component sector and should originate only from one
loop.\footnote{Recently it was shown \cite{KM2} that unlike ${\cal
N}=4$ SYM theory, $F^4$ term can appear in two-loop effective action
for ${\cal N}=2$ superconformal field theories.} It means, that
two-loop correction to classical action should begin at least with
$F^6$ term.  The corresponding calculations have been done in
\cite{BPT}. The main motivation of the work \cite{BPT} was to study the
relation between ${\cal N}=4$ SYM effective action and $D3$-brane
interactions in superstring theory. In static limit this interaction is
described by Born-Infeld action which can be presented as a series in
powers of Abelian strength $F_{mn}$ (see the review \cite{T}). It was
proved in \cite{BPT} that in 't Hooft limit the coefficient at two-loop
$F^6$ term in ${\cal N}=4$ SYM effective action exactly coincides with
the coefficient at $F^6$ term in expansion of Born-Infeld action.
Coincidence of the coefficients at $F^4$ terms was earlier demonstrated
in \cite{BKT}.  These results confirm the conjecture that $D3$-brane
interactions can be completely described in terms of ${\cal N}=4$ SYM
theory. We discuss further the basic elements of calculating the
leading two-loop contribution to effective action in 't Hooft limit
using the harmonic superspace formulation of ${\cal N}=4$ SYM theory.

In ${\cal N}=2$ harmonic superspace the ${\cal N}=4$ SYM
theory is formulated in terms of ${\cal N}=2$ gauge superfield and
hypermultiplet superfield. We consider ${\cal N}=4$ SYM theory with
gauge group $SU(N+1)$ spontaneusly broken down to $SU(N)$ x $U(1)$ and
study the effective action depending on the background superfields
${\cal W}, \bar{\cal W}$ belonging to Abelian factor $U(1)$. The
effective action is calculated on the base of ${\cal N}=2$
supersymmetric background field method \cite{BBKO} in large $N$ limit.
Two-loop contibution to effective action is given by the following
harmonic supergraphs
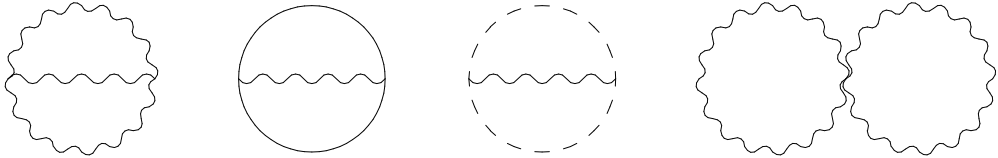
\vspace{6mm}

\Lengthunit=1.95cm
\mov(1.25,0){\wavecirc(1.0)\mov(-0.5,0){\wavelin(1.0,0)}
\mov(1.5,0){\Circle(1.0)}\mov(2.0,0){\wavelin(-1.0,0)}
\mov(3.0,0){\dashcirc(1.0)}\mov(3.5,0){\wavelin(-1.0,0)}
\mov(4.5,0){\wavecirc(1)\mov(1.0,0){\wavecirc(1)}}
}
\vspace{6mm}

\noindent
where the wavy, solid and dashed lines stand for the propagators of
${\cal N}=2$ gauge, hypermultiplet and ghost superfields respectively.
All propagators depend on Abelian on-shell background superfields
${\cal W}, \bar{\cal W}$. Explicit forms of the propagators are given
in \cite{BBKO}, \cite{BPT}. In the above supergraphs we expand each
propagator in ${\cal W}, \bar{\cal W}$ and
their spinor derivatives and look for the leading low-energy
contribution.  One can show that the expected manifestly ${\cal N}=2$
supersymmetric $F^6$ term is generated by the terms in effective action
containing the fourth order in $D^{i}_{\alpha}{\cal W}$ and second
order in ${\bar {D}}_{i{\dot{\alpha}}}\bar{\cal W}$ (plus the conjugate
terms). It was proved in \cite{BPT} that the supergraphs including the
hypermultiplet and ghost lines and supergraph with quartic gauge
superfield vertex do not contribute to effective action in leading
low-energy approximation. Thus, we have to study the only
supergraph with two cubic gauge superfield vertices. Calculation of
this supergraph is done using the standard $D$-algebra manipulations.
The final result for two-loop correction
$\Gamma_{2}[{\cal W},\bar{\cal W}]$ to effective action has the form
\begin{eqnarray}
\Gamma_{2}[{\cal W},\bar{\cal W}] &=& c_{2}g^{2}N^{2} \int {\rm
d}^{12}z {1\over {\bar{{\cal W}^2}}}{\rm ln}\frac{{\cal W}}{{\Lambda}}
D^{4}{\rm ln}\frac{{\cal W}}{{\Lambda}} + h.c. \label{25}
\end{eqnarray}
where the $c_{2} = {1\over {24(4\pi)^{4}}}$.\footnote{The coefficient
$c_{2}$ was given in \cite{BPT} in terms of coupling $g_{YM}^{2} =
{1\over {2}}g^{2}$ where $g$ is the gauge coupling we use here.} Eq
(\ref{25}) determines the leading two-loop low-energy contribution to
effective action The coefficient $c_{2}$ exactly corresponds to the
coefficient at $F^6$ term in expansion of Born-Infeld action
\cite{BPT}.

\section{Summary}
We have considered the current state of problem of effective action in
${\cal N}=4$ SYM theory. The main results are formulated as
follows:

{\bf 1.} Low-energy effective action depending on all fields of ${\cal
N}=4$ vector multiplet is exactly found.

{\bf 2.} One-loop effective action depending on all fields of
${\cal N}=4$ vector multiplet is exactly found for the supersymmeric
background corresponding to constant Abelian strength $F_{mn}$.

{\bf 3.} Leading two-loop contribution to low-energy effective action
containing $F^6$ term in bosonic component sector is found.

{\bf 4.} The coeffcients at classical $F^{2}$ term, one-loop
$F^{4}$ term and two-loop $F^{6}$ term in effective action exactly
coincide with the corresponding coefficients in expansion of
Born-Infeld action.

\section{Acknowledgements}
Author is very grateful to INTAS grant, INTAS-03-51-6346, RFBR grant,
project No 03-02-16193, LSS grant, project No 1252.2003.2 and DFG
grant, project No 436 RUS 113/669/0-2 for partial support.

\end{document}